\documentclass[useAMS,usenatbib,onecolumn,usegraphicx]{mn2e}

\newcommand{\lsim}{\, \, \raisebox{-0.8ex}{$\stackrel{\textstyle <}{\sim}$ }}

\newcommand{\beq}{\begin{equation}}
\newcommand{\eeq}{\end{equation}}
\newcommand{\beqar}{\begin{eqnarray}}
\newcommand{\eeqar}{\end{eqnarray}}

\title[Probing nuclear bubble structure]  
{Probing nuclear bubble structure via neutron star asteroseismology}
\author[H. Sotani, K. Iida, \& K. Oyamatsu]
{Hajime Sotani$^1$ \thanks{E-mail:sotani@yukawa.kyoto-u.ac.jp},
Kei Iida$^2$, and
Kazuhiro Oyamatsu$^3$
\\
$^1$Division of Theoretical Astronomy, National Astronomical Observatory of Japan, 2-21-1 Osawa, Mitaka, Tokyo 181-8588, Japan\\
$^2$Department of Natural Science, Kochi University, 2-5-1 Akebono-cho, Kochi 780-8520, Japan\\
$^3$Department of Human Informatics, Aichi Shukutoku University, 2-9 Katahira, Nagakute, Aichi 480-1197, Japan}

\begin{document}
\maketitle
\label{firstpage}

\begin{abstract}
We consider torsional oscillations that are trapped in a layer of spherical-hole 
(bubble) nuclear structure, which is expected to occur in the deepest region of 
the inner crust of a neutron star.  Because this layer intervenes between the phase
of slab nuclei and the outer core of uniform nuclear matter, torsional oscillations 
in the bubble phase can be excited separately from usual crustal torsional oscillations.
We find from eigenmode analyses for various models of the equation of state of
uniform nuclear matter that the fundamental frequencies of such oscillations are almost 
independent of the incompressibility of symmetric nuclear matter, but strongly depend 
on the slope parameter of the nuclear symmetry energy $L$.  Although the frequencies 
are also sensitive to the entrainment effect, i.e., what portion of nucleons outside
bubbles contribute to the oscillations, by having such a portion fixed, we can 
successfully fit the calculated fundamental frequencies of torsional oscillations 
in the bubble phase inside a star of specific mass and radius as a function of $L$.  
By comparing the resultant fitting formula to the frequencies of quasi-periodic 
oscillations (QPOs) observed from the soft-gamma repeaters, we find that each of 
the observed low-frequency QPOs can be identified either as a torsional 
oscillation in the bubble phase or as a usual crustal oscillation, given 
generally accepted values of $L$ for all the stellar models considered here. 
\end{abstract}


\section{Introduction}
\label{sec:I}

Neutron star crusts, which are composed of inhomogeneous neutron-rich nuclear 
matter embedded in a roughly uniform neutralizing background of electrons, can be 
unique laboratories to provide us with simultaneous manifestations of superfluids,
liquid crystals, and solids.  This is because nuclear matter at subnuclear
densities and sufficiently low temperatures can exhibit various mixed phases of
a liquid composed of protons and neutrons and a neutron gas as a result of the 
combined effect of the tensor, isoscalar part of the nuclear force and the Coulomb 
interaction, 
while keeping both the liquid and the gas in a superfluid state thanks to the 
central, isovector part of the nuclear force \citep{PR1995}.  The crustal region, 
which is located in the outer part of a neutron star, is observationally more 
relevant than the inner part, namely, the core region, because of the closer 
distance to the star's magnetosphere and surface, from which electromagnetic 
emission occurs.

From the viewpoint of condensed matter physics, however, the deeper, the more 
interesting.  In fact, roughly spherical nuclei (liquid part), which are predicted 
to form a body-centered cubic (bcc) lattice at relatively low densities, are 
considered to fuse into rod-like nuclei in a gas of neutrons when the spherical 
nuclei become so closely packed as to be almost unstable with respect to 
quadrupolar deformations. As the density 
increases further, it is expected that the shape of the liquid part 
in the crust changes from spherical to cylindrical (rod), slab, 
cylindrical-hole (tube), and spherical-hole (bubble) structures until matter 
becomes uniform \citep{LRP1993,O1993}.  The rod, slab, tube, and bubble 
structures  are often referred to as nuclear 
pasta.  Since the spherical and bubble phases are solids while the cylindrical, 
slab, and cylindrical-hole phases are liquid crystals, possible observations
of global free oscillations from neutron stars could be useful for  
obtaining information about elastic and superfluid properties of neutron star 
interiors \citep{PA2012}.   This technique is known as asteroseismology, 
which is essentially the same as seismology in the case of the Earth 
and helioseismology in the case of the Sun.  In fact, it has been 
suggested that the neutron star's properties such as the mass and 
radius, the equation of state (EOS) of matter therein, and the magnetic 
properties would be possible to obtain via the spectra of the star's 
oscillations (see, e.g., \cite{VH1995}).

Neutron star asteroseimology is unique in the sense that in addition to 
electromagnetic waves, gravitational waves radiating from the star are 
expected to provide us with information about the star's global 
oscillations \citep{AK1996,STM2001,SKH2004,SYMT2011,DGKK2013}.  Direct 
gravitational wave detections from neutron stars, which have yet to be done, 
would be highly promising in the near future.

Meanwhile, there are X-ray observational evidences for neutron star 
oscillations.  In fact, quasi-periodic oscillations (QPOs) were 
discovered in the X-ray afterglow of giant flares from soft-gamma 
repeaters (SGRs) \citep{I2005,SW2005,SW2006,QPO2}, which are supposed to be
strongly magnetized neutron stars \citep{K1998,H1999}.  Although there are 
still many uncertainties in understanding of the mechanism of the giant 
flares and the subsequent QPOs, it is generally accepted that the QPOs 
arise from global oscillations of the neutron stars.  The observed QPO 
frequencies are in the range of tens Hz up to kHz, while typical frequencies 
of neutron star acoustic oscillations are around kHz \citep{VH1995}.
Particularly, identification of the QPO frequencies lower than $\sim 100$ Hz 
is not straightforward but could significantly constrain the possible 
origin of the QPOs.  Basically, candidates for the corresponding global 
oscillations are crustal torsional oscillations, magnetic oscillations,
and coupled oscillations between these two.

Global magnetic oscillations in neutron stars depend crucially  
on the magnetic field strength and structure therein \citep{GCFMS2013}, 
but those are still poorly known, particularly in superconducting 
materials, as well as the EOS for matter in the core.  In order to avoid such 
uncertainties, in this paper we simply consider the observed low-lying 
QPOs as crustal torsional oscillations.  In fact, within such identifications, 
one can obtain information about the crustal properties by fitting 
the calculated eigenfrequencies to the observed QPO frequencies
\citep{SA2007,SW2009,GNJL2011,SNIO2012,SNIO2013a,SNIO2013b,S2014,S2016,SIO2016}.
Because of the success in accurately reproducing all the QPO frequencies, this kind 
of approach might well play the role of a canonical model for the low-lying QPOs 
from SGRs.  There is nevertheless a serious caveat: To obtain the shear modulus
that is consistent with the QPO frequencies, one requires a significantly
large value of the parameter $L$ that characterizes the density dependence of 
the symmetry energy of nuclear matter, as compared with what various nuclear 
observables suggest.

So far, several calculations of the eigenfrequencies of crustal torsional 
oscillations have been done by including the effect of superfluidity, 
but the effect of the possible existence of the 
pasta structure has been neglected in most of such calculations; an artificial 
shear modulus has been at most taken into consideration for the pasta phases 
\citep{S2011,PP2016}.  Since the crystalline structure in the bubble phase is 
presumably the same as that in the low density region composed of 
spherical nuclei, however, one can likewise calculate the 
eigenfrequencies of the torsional oscillations in the bubble phase.  As we 
shall see, furthermore, the smectic-A liquid-crystalline properties in the 
phase with slab-shaped nuclei \citep{PP1998} do not allow torsional shear 
oscillations to occur in linear analysis, which leads to the conclusion 
that the torsional oscillations in the bubble phase can be excited separately 
from those in the low density regime.  Bearing this in mind, we search for 
a better fitting to the observed low-lying QPO frequencies while keeping the value of
$L$ reasonable.

In Sec.\ \ref{sec:II} the equilibrium configuration of a neutron star crust is 
constructed.  Section \ref{sec:III} is devoted to eigenmode analyses of torsional 
shear oscillations within the bubble phase.  The resultant eigenfrequencies are
compared with the observed QPO frequencies in Sec.\ \ref{sec:IV}.
Concluding remarks are given in Sec.\ \ref{sec:V}.
We use units in which $c=G=1$, where $c$ and $G$ 
denote the speed of light and the gravitational constant, respectively.

\section{Crust in equilibrium}
\label{sec:II}

We start with description of the equilibrium configuration of a neutron star 
crust.  In this description, we need the EOS of equilibrated crustal matter.  For 
simplicity, we assume that the temperature of the matter is zero.  This is a 
very good approximation in analyzing the crust's equilibrium configuration and 
eigenfrequencies of its torsional oscillations, but is not necessarily so in 
describing damping of such oscillations, which is beyond the scope of the 
present analysis.

As discussed in \cite{L1981}, the bulk energy per nucleon of uniform 
nuclear matter at zero temperature can be generally expanded as a 
function of baryon number density $n_{\rm b}$ and neutron excess $\alpha$:
\begin{equation}
  w = w_0  + \frac{K_0}{18n_0^2}(n_{\rm b}-n_0)^2 + \left[S_0 
      + \frac{L}{3n_0}(n_{\rm b}-n_0)\right]\alpha^2.
  \label{eq:w}
\end{equation}
Here $w_0$, $n_0$, and $K_0$ denote the saturation energy, saturation density, and 
incompressibility of symmetric nuclear matter ($\alpha=0$), while $S_0$ and $L$ are 
associated with the density dependent symmetry energy $S(n_{\rm b})$, i.e., 
$S_0\equiv S(n_0)$ and $L\equiv 3n_0(dS/dn_{\rm b})_{n_{\rm b}=n_0}$.  Since $w_0$, $n_0$, 
and $S_0$ characterize the saturated, nearly symmetric nuclear matter, 
these three parameters are relatively well constrained from empirical data for 
masses and radii of stable nuclei as compared to the remaining parameters 
$L$ and $K_0$, which characterize the way the energy changes as the baryon 
density changes from the saturation point of symmetric nuclear matter.  
With increasing $\alpha$, the saturation density $n_s$ and energy $w_s$ 
changes from $n_0$ and $w_0$ as
\begin{equation}
  n_s = n_0-\frac{3n_0 L}{K_0}\alpha^2,
  \label{ns}
\end{equation}
\begin{equation}
  w_s = w_0+S_0\alpha^2.
  \label{ws}
\end{equation}

Two of the authors \citep{OI2003}
have derived a phenomenological model for the EOS of uniform nuclear 
matter in such a way that the bulk energy of nuclear matter reproduces 
Eq.\ (\ref{eq:w}) in the limit of $n_{\rm b}\to n_0$ and $\alpha\to 0$.  Within a 
simplified version of the extended Thomas-Fermi theory for a nucleus of 
mass number $A$, the most relevant values of $w_0$, $n_0$, and $S_0$ were obtained 
by fitting the calculated mass excess, charge radius, and charge number of 
stable nuclei to the empirical data for given values of $y\equiv -K_0S_0/(3n_0L)$ 
and $K_0$.  To obtain the equilibrium nuclear shape and size in neutral matter 
in the crust, furthermore, \cite{OI2007} have derived the optimal energy density 
and nucleon distribution as a function of $n_{\rm b}$ within a Wigner-Seitz 
approximation by taking into account the presence of a gas of dripped 
neutrons and a uniform electron gas, as well as the five nuclear shapes 
(sphere, cylinder, slab, tube, bubble).   The EOS parameter sets adopted in this 
paper are shown in Table \ref{tab:SH-density}.

We turn to the equilibrium phase transitions between various nuclear shapes.  
For description of such phase transitions for various sets of the EOS parameters 
$L$ and $K_0$, as in \cite{OI2007}, we obtain the optimal energy densities of the 
five liquid-gas mixed phases and the homogeneous phase at given baryon density 
and then identify the phase with minimal energy density, i.e., the equilibrium phase.  
Note that all the phase transitions considered here are of first order.  In a star, 
where the pressure increases continuously with depth, the baryon density is
discontinuous at the transition points.  Since such discontinuities are negligibly 
small, however, we will regard it as continuous in calculating the star's structure.

\begin{figure*}
\begin{center}
\begin{tabular}{cc}
\includegraphics[scale=0.5]{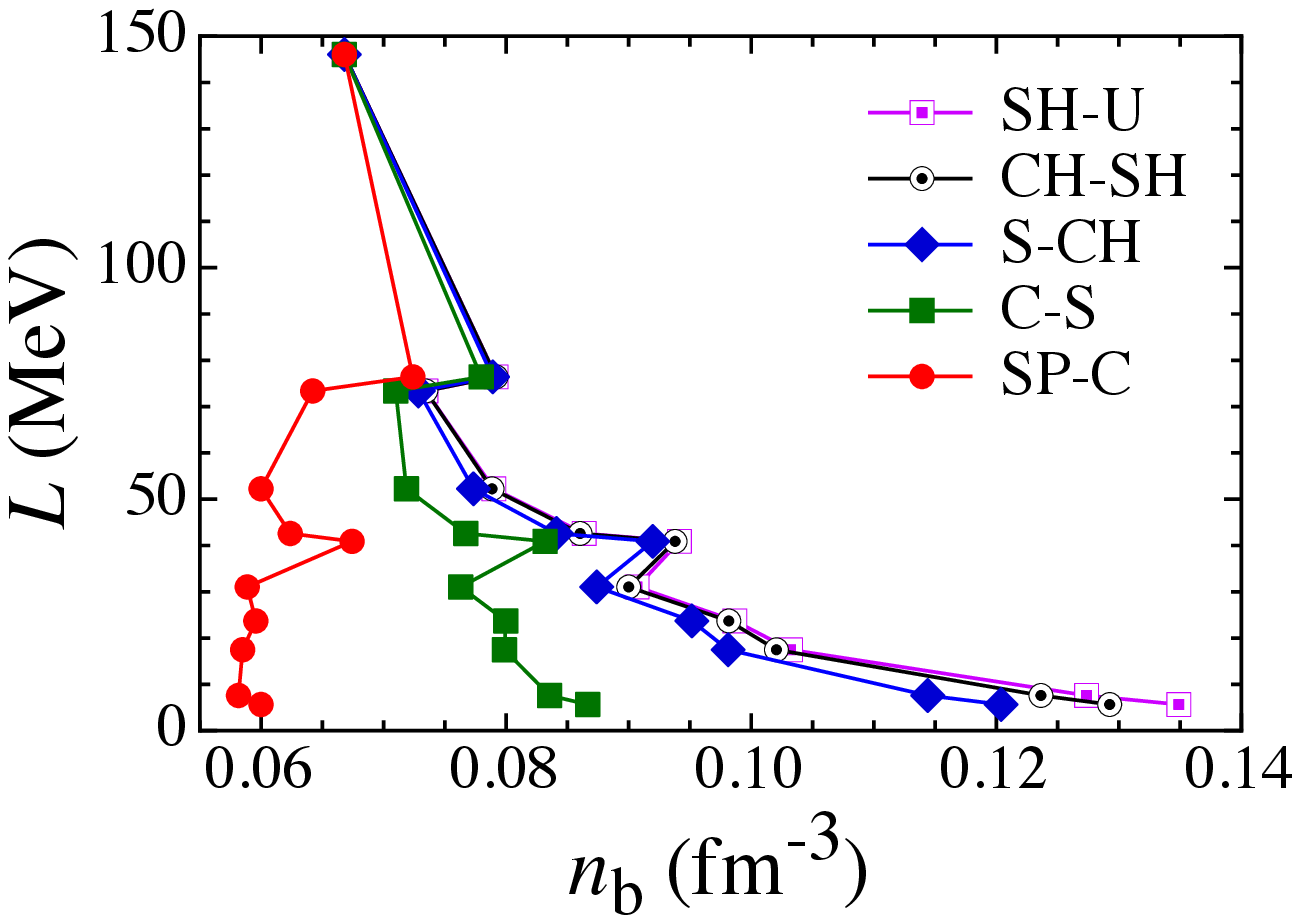} &
\includegraphics[scale=0.5]{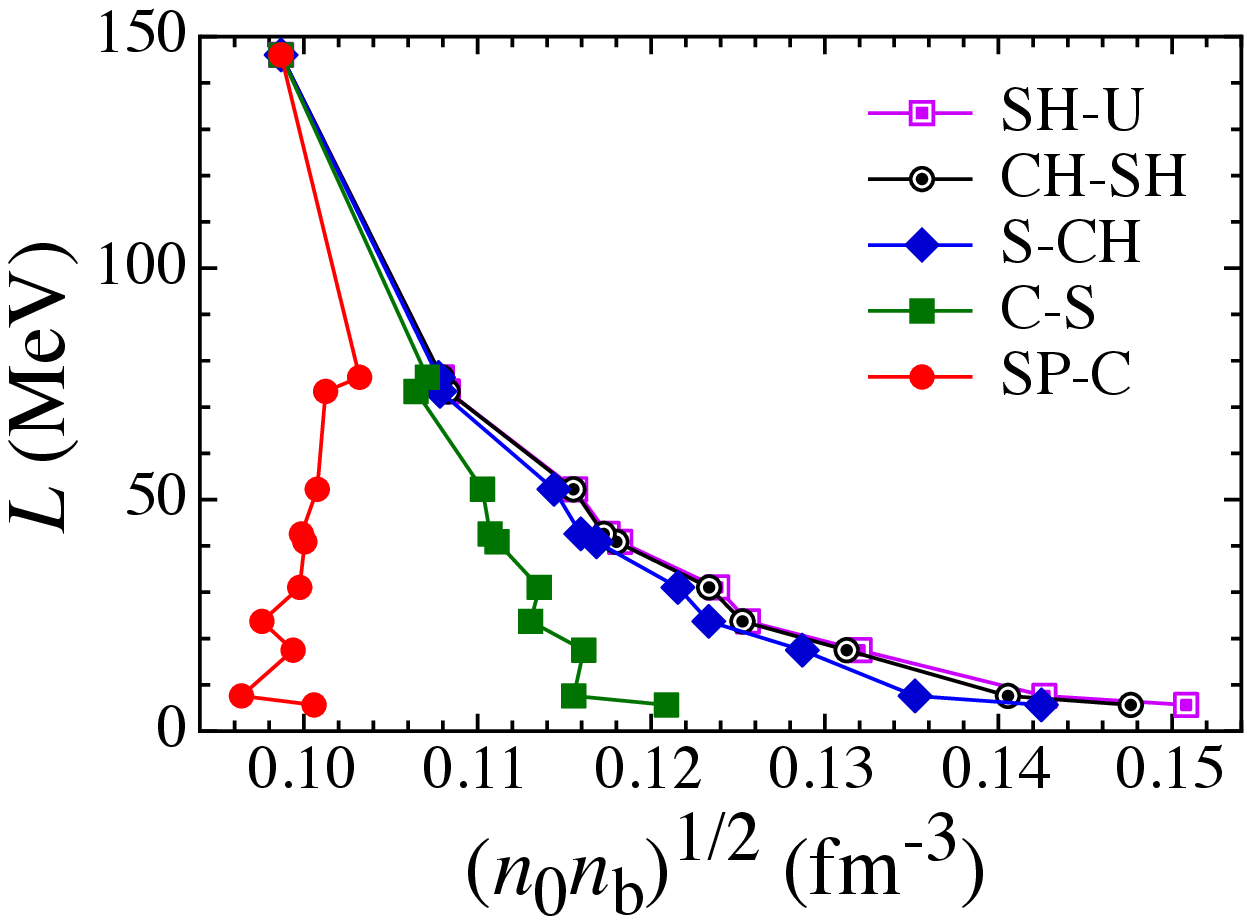}
\end{tabular}
\end{center}
\caption{
(Color online) The baryon number density $n_{\rm b}$ (left) and its combination with $n_0$, 
$(n_0 n_{\rm b})^{1/2}$ (right), at the structural phase transitions, as plotted
for eleven EOS models which are classified by the value of $L$.  Here, the 
circles, squares, diamonds, double circles, and double squares represent the 
transition densities from spherical to cylindrical nuclei (SP-C), from 
cylindrical to slablike nuclei (C-S), from slablike to cylindrical-hole nuclei (S-CH), 
from cylindrical-hole to spherical-hole (bubble) nuclei (CH-SH), and from spherical-hole 
nuclei to uniform matter (SH-U), respectively.  The data are extracted from 
Oyamatsu \& Iida (2007).
}
\label{fig:nb-L}
\end{figure*}

In Fig.\ \ref{fig:nb-L}, for each EOS parameter set ($K_0$, $L$), we display 
the resultant baryon density at the phase transitions from spherical to 
cylindrical nuclei (SP-C), from cylindrical to slablike nuclei (C-S), from slablike 
to cylindrical-hole nuclei (S-CH), from cylindrical-hole to spherical-hole (bubble) 
nuclei (CH-SH), and from spherical-hole nuclei to uniform matter (SH-U), respectively. 
In this figure (left panel), the $L$ dependence of the transition densities 
is remarkable in the sense that all the transition densities seemingly converge to 
$\sim0.07$ fm$^{-3}$ with increment in $L$.  This is because a larger $L$, or, 
equivalently, a smaller symmetry energy at subnuclear densities, acts to decrease 
the density of a liquid part (see Eq.\ (\ref{ns})) and simultaneously increase 
the density of a gas of dripped neutrons, as can be clearly seen in Fig.\ 6 of 
\cite{OI2007}.  Moreover, the $K_0$ dependence of the transition densities is 
implicit in the fluctuating pattern of each transition line: The larger $K_0$, 
the larger transition density.  Instead of these transition baryon densities, 
we plot, in right panel, $(n_0n_{\rm b})^{1/2}$ where we set $n_{\rm b}$ to the 
transition densities.  Apparently, $(n_0n_{\rm b})^{1/2}$ has only a weak dependence 
on $K_0$.  This is because the tendency that $n_0$ decreases with increasing $K_0$, 
which comes mainly from the fitting to empirical charge radii of stable nuclei 
\citep{OI2003}, plays a role in counteracting the $K_0$ dependence of the 
transition densities.  Incidentally, the latter dependence stems partly from the 
conclusion of a liquid-drop model \citep{OHY1984} that each shape transition occurs 
when the volume fraction of a liquid part reaches a critical constant value and 
partly from the fact that at large neutron excess, $K_0$ affects the relation 
between the volume fraction and the baryon density 
by increasing the density of a liquid part (see Eq.\ (\ref{ns})).

Additionally, in Table \ref{tab:SH-density}, we list the values of the phase 
transition density from spherical to cylindrical nuclei, from cylindrical-hole to 
spherical-hole nuclei, and from spherical-hole nuclei to uniform matter.  We remark 
that all the pasta structures are predicted to appear for all the 
EOS parameter sets shown in this table except the cases of $(K_0,L)=(360,76.4)$ 
and $(360,146.1)$ in MeV.  In practice, no pasta structures appear 
for the case of $(K_0,L)=(360,146.1)$, i.e., spherical nuclei transform 
directly to uniform matter at $n_{\rm b}=0.06680$ fm$^{-3}$, while only the bubble 
structure is absent for $(K_0,L)=(360,76.4)$, i.e., cylindrical-hole nuclei 
transform to uniform matter at $n_{\rm b}=0.07918$ fm$^{-3}$.  Thus, the maximum 
value of $L$ for the bubble structure to appear in neutron stars is 
predicted to be $L\simeq 75$ MeV.\footnote{The maximum value of $L$ for
any pasta structures to appear, which is of order 100 MeV, has already been 
discussed in \cite{OI2007} in terms of fission-like instability of spherical
nuclei as well as proton clustering instability in uniform nuclear matter at 
subnuclear densities.}  Since we focus on torsional oscillations confined
in the bubble phase in this work, we shall consider only the 
cases with $L\lsim75$ MeV.

\begin{table}
\centering
\begin{minipage}{100mm}
\caption{
The SP-C, CH-SH, and SH-U transition densities obtained for each EOS model, 
which is characterized by $K_0$ and $L$. The asterisk at the value of $K_0$ 
denotes the EOS model by which the spherical-hole phase is not predicted 
to appear.  In this case, the SH-U transition density should read the density
at which the system melts into uniform matter.
}
\begin{tabular}{cc|cccc}
\hline\hline
  $K_0$ (MeV) & $L$ (MeV) & SP-C (fm$^{-3}$) & CH-SH (fm$^{-3}$) & SH-U (fm$^{-3}$)  \\
\hline
  180 & 5.7   & 0.06000 & 0.12925   & 0.13489     \\  
  180 & 17.5 & 0.05849 & 0.10206  &  0.10321     \\  
  180 & 31.0 & 0.05887 & 0.09000   & 0.09068     \\  
  180 & 52.2 & 0.06000 & 0.07885   & 0.07899     \\  
  230 & 7.6   & 0.05816 & 0.12364   & 0.12736     \\  
  230 & 23.7 & 0.05957 & 0.09817  &  0.09866     \\  
  230 & 42.6 & 0.06238 & 0.08604   & 0.08637     \\  
  230 & 73.4 & 0.06421 & 0.07344   & 0.07345     \\  
  360 & 40.9 & 0.06743 & 0.09379  &  0.09414     \\  
  $^*$360 & 76.4   & 0.07239 & ---   & 0.07918     \\  
  $^*$360 & 146.1 & 0.06680 & ---   & 0.06680     \\  
\hline\hline
\end{tabular}
\label{tab:SH-density}
\end{minipage}
\end{table}

The equilibrium configuration of the crust of a spherically symmetric 
neutron star is constructed by integrating the Tolman-Oppenheimer-Volkoff 
(TOV) equations inward from the star's surface down to the 
crust-core boundary in combination with the crust EOS \citep{IS1997}, 
in such a way that we do not have to use the core EOS, which is uncertain, 
explicitly.  We remark that the crust-core boundary is set to the 
position where the phase transition occurs from spherical-hole nuclei 
into uniform matter in the present analysis, while, in the previous studies 
\citep{SNIO2012,SNIO2013a,SNIO2013b,SIO2016,S2016}, being simply set to 
the position where the phase transition occurs from spherical nuclei 
into cylindrical nuclei or uniform matter, depending on whether or not 
the phase with cylindrical nuclei can be energetically favorable.  Under spherical 
symmetry, the metric can be obtained in terms of the spherical polar 
coordinates $r$, $\theta$, and $\phi$ as
\begin{equation}
 ds^2 = -{\rm e}^{2\Phi}dt^2 + {\rm e}^{2\Lambda}dr^2 + r^2 d\theta^2 
+ r^2\sin^2\theta\, d\phi^2,  \label{metric}
\end{equation}
where $\Phi$ and $\Lambda$ are the metric functions that depend only on $r$.
The mass function $m(r)$ is associated with $\Lambda$ via 
$\exp(-2\Lambda)=1-2m/r$.

\begin{figure*}
\begin{center}
\begin{tabular}{cc}
\includegraphics[scale=0.5]{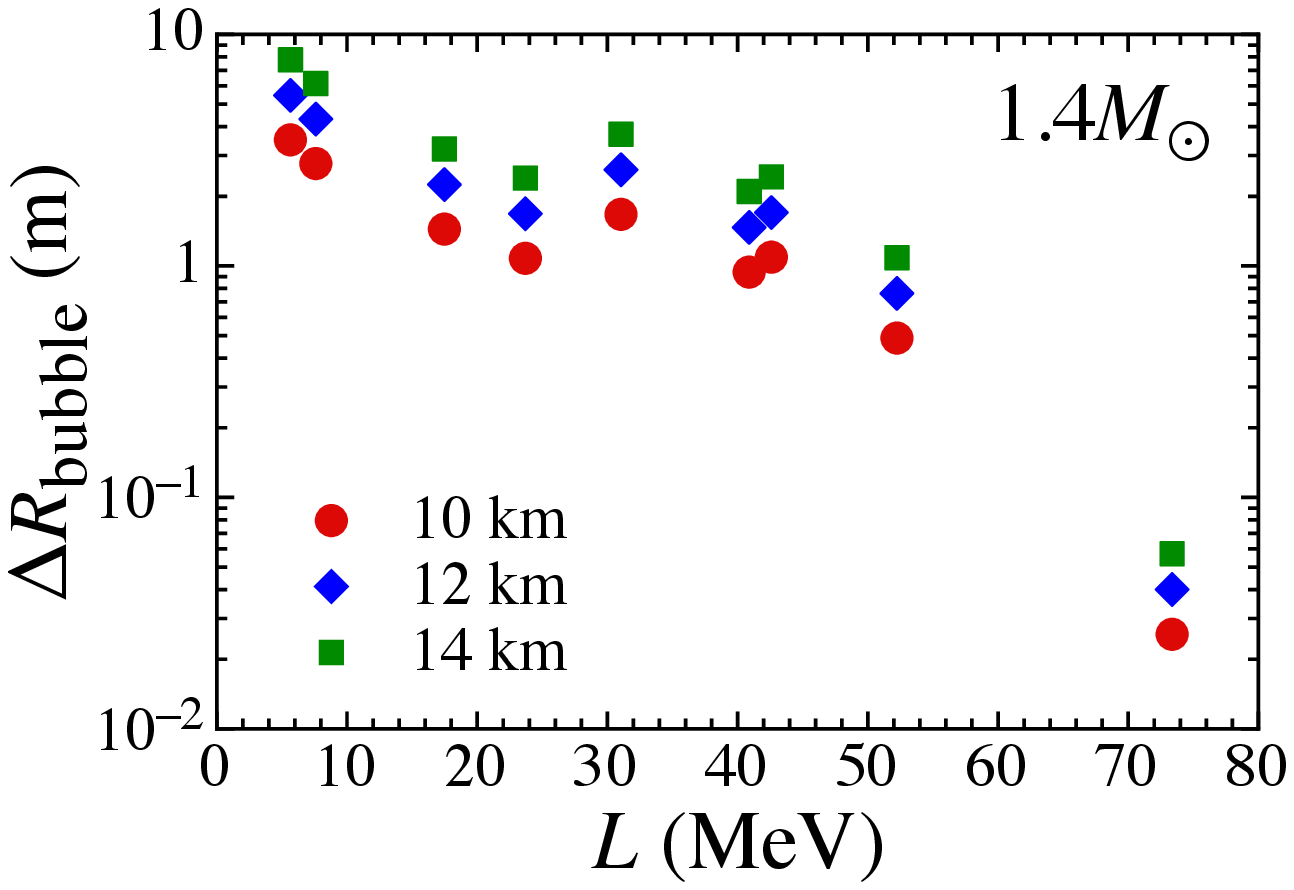} &
\includegraphics[scale=0.5]{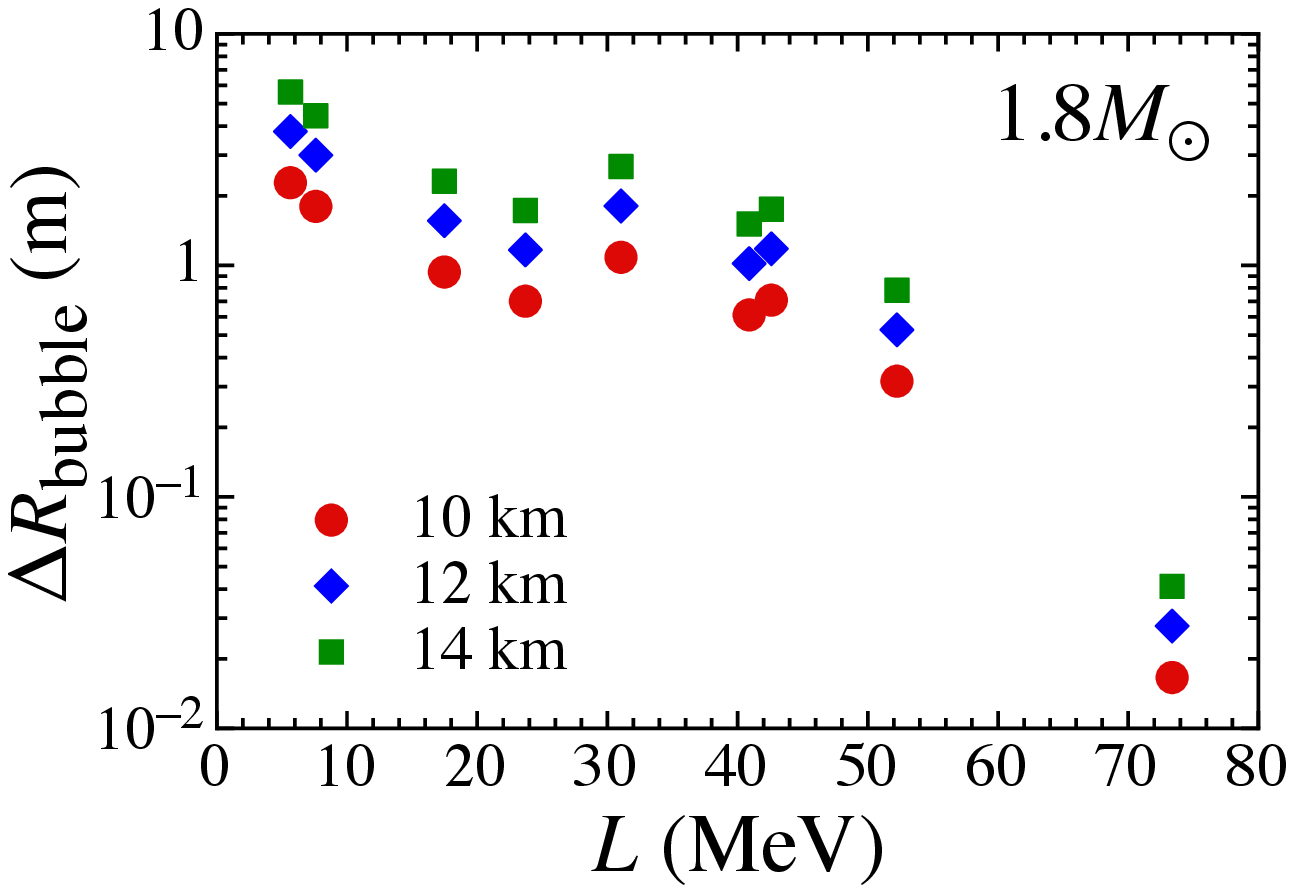}
\end{tabular}
\end{center}
\caption{
(Color online) Thickness of the bubble phase in the crust of a neutron star.  The left 
and right panels correspond to the cases of $M=1.4M_\odot$ and $1.8M_\odot$, 
respectively, while the circles, diamonds, and squares denote the 
cases of $R=10$, 12, and 14 km.
}
\label{fig:dr}
\end{figure*}

According to the solutions to the TOV equations, the thickness of the 
bubble phase for typical neutron star models with mass $M$ and radius $R$ depends 
sensitively on $L$ and is at most $\sim 10$ m.  In Fig.\ \ref{fig:dr}, the thickness 
of the bubble phase is shown for the stellar models with $M=1.4M_\odot$, $1.8M_\odot$ 
and $R=10$, 12, 14 km.
We remark that the spherical layer of bubbles is located only within $1.5$ km from the 
star's surface for all the stellar models considered here.  It is not the thickness but the
radius of this layer that is essential to the fundamental frequencies of torsional 
oscillations trapped therein. The overtone frequencies are sensitive to the thickness,
but are high enough to be beyond the scope of this paper.

The shear modulus of crustal matter is one of the most important properties 
in describing crustal torsional oscillations.  For example, the shear modulus 
of a bcc lattice of spherical nuclei has been calculated 
as a function of the charge number $Z$, the number density of nuclei $n_i$, 
and the Wigner-Seitz radius $a$ as \citep{OI1990,SHOII1991}
\begin{equation}
   \mu = 0.1194\frac{n_i (Ze)^2}{a}, \label{eq:shear}
\end{equation}
by assuming that each nucleus is a point particle.  Although modifications 
of the shear modulus by electron screening and polycrystalline nature 
have also been considered by \cite{KP2013,KP2015}, we adopt the traditional 
formula for the shear modulus [Eq.\ (\ref{eq:shear})] for simplicity
in the present analysis.   Since the crystalline structure in the bubble phase is 
presumably the same as that in the phase composed of spherical nuclei, we will 
consider torsional oscillations in the bubble phase by using the shear modulus 
given by Eq.\ (\ref{eq:shear})  with appropriate replacements.  For the bubble phase,
we reinterpret $n_i$ as the number density of bubbles and, as in \cite{WI2003},
replace $Z$ with the effective charge number $Z_{\rm bubble}$ of a bubble.  Here, 
the effective charge number is given by $Z_{\rm bubble}=n_Q V_{\rm bubble}$, where 
$V_{\rm bubble}$ denotes the volume of the bubble, and $n_Q$ is the effective charge 
number density inside the bubble.  $n_Q$ can be calculated as 
$n_Q=-n_e-(n_p-n_e)=-n_p$ with the number density of protons outside the bubble, 
$n_p$, and the number density of a uniform electron gas, $n_e$,  because,  
in the bubble phase, the background (outside the bubble) charge number density is 
$n_p-n_e$, while the charge number density inside the bubble is $-n_e$.

The elastic properties in the pasta phases of cylindrical and slab nuclei have also 
been discussed in terms of liquid crystals \citep{PP1998}.  In fact, these phases
can be regarded as a columnar phase and a smectic A, respectively.  For these
liquid crystals, the elastic properties and the propagating modes are well known
\citep{LL1986,dGP1993}.  \cite{PP1998} utilized an incompressible liquid-drop model for 
pasta nuclei to derive the relation between the elastic constants involved and the 
Coulomb energy density.  We remark that the elastic properties in the cylindrical-hole 
phase is essentially the same as that in the cylindrical phase.  The important 
aspect of global torsional oscillations in the crust is that 
a restoring force due to the shear stress is responsible for the propagation of 
the oscillations.  To linear order in displacements of the pasta nuclei, there is
no such restoring force in the slab phase,\footnote{Even in the incompressible 
limit, there is a propagating transverse mode, often referred to as a ``second sound'' 
\citep{LL1986,dGP1993}.  This mode involves spatially varying interlayer compression,
which in turn couples with spatially varying fluid pressure.  A similar kind of 
``second sound" wave can occur in the columnar phase.  All of these modes are beyond 
the scope of the present analysis.} while nonzero shear modulus in the cylindrical-hole 
phase plays a role in propagation of a torsional shear wave, often referred to as a 
``third sound.''   Thus, the torsional oscillations that are excited within the
cylindrical-hole and bubble phases are expected to be separable from those 
within the spherical and cylindrical phases, although there could be nonlinear
coupling between these two.  For simplicity, in the next section, we will consider the 
torsional oscillations that propagate only in the bubble phase;  a ``third 
sound'' in the cylindrical hole phase and its connection with the torsional oscillations 
in the bubble and cylindrical phases will be allowed for elsewhere.

\section{Bubble torsional oscillations}
\label{sec:III}

Let us now calculate the eigenfrequencies of torsional oscillations in the
bubble phase that is embedded in the spherically symmetric equilibrium  
configuration of a neutron star crust as constructed in the previous section.
We start with the case in which all the matter components participate in the the
oscillations.  Since torsional oscillations are incompressible, i.e., the 
oscillations do not involve the density variation, one can determine their frequencies 
with high accuracy even within the relativistic Cowling approximation 
in which the metric is fixed during the oscillations.  The perturbation equation 
that governs the torsional oscillations can be derived from the linearized 
equation of motion as \citep{ST1983}
\begin{equation}
 {\cal Y}'' + \left[\left(\frac{4}{r}+\Phi'-\Lambda'\right)+\frac{\mu'}{\mu}\right]{\cal Y}'  
      + \left[\frac{H}{\mu}\omega^2{\rm e}^{-2\Phi}-\frac{(\ell+2)(\ell-1)}{r^2}\right]
        {\rm e}^{2\Lambda}{\cal Y} = 0,
 \label{eq:perturbation}
\end{equation}
where ${\cal Y}$ denotes the Lagrangian displacement in the $\phi$ direction, $H$ is 
the enthalpy density defined as $H\equiv p + \varepsilon$ with the pressure $p$ and 
energy density $\varepsilon$, and $\ell$ is the angular index. ${\cal Y}$ is 
associated with the $\phi$ component of the perturbed four velocity as $\delta u^{\phi} 
= {\rm e}^{-\Phi}\partial_t {\cal Y}(t,r)(\sin\theta)^{-1}\partial_\theta P_{\ell}(\cos\theta)$, 
where $P_{\ell}(\cos\theta)$ is the $\ell$-th order Legendre polynomial.  As mentioned 
in the previous section, we assume the situation in which torsional oscillations
occur solely in the bubble phase.  In terms of the boundary conditions,  this
situation conforms to the zero-traction conditions, i.e., ${\cal Y}'=0$ at the 
inner and outer boundaries of the bubble phase.

It is well known that the frequencies of torsional oscillations are proportional 
to the shear velocity defined as $v_s=(\mu/H)^{1/2}$ \citep{HC1980}.  Thus, not 
only $\mu$, but also the enthalpy density plays a role in determining the 
frequencies of torsional oscillations.  As in the case of spherical nuclei in a 
neutron superfluid \citep{SNIO2013a}, we here consider the reduction of the enthalpy 
density by superfluidity of nuclear matter outside the bubbles.  If the whole nuclear 
matter behaves as a superfluid and only a neutron gas inside the bubbles participate in
the oscillations, then, the effective enthalpy density $\tilde{H}$ that contributes to 
the oscillations would be minimal.  In a real world, however, a portion of nuclear
matter outside the bubbles comove nondissipatively with the bubbles by undergoing 
Bragg scattering off the bcc lattice of the bubbles.  This effect, often denoted by 
the entrainment effect, was originally considered for a neutron gas dripped out of the
spherical nuclei \citep{Chamel2012}, and hence the effective enthalpy density 
$\tilde{H}$ could be quantitatively estimated from similar band calculations.  Instead 
of performing such calculations, we mainly analyze the two extreme cases in 
which the effective enthalpy density is maximal, i.e., $\tilde{H}=H$, and minimal,
corresponding to the minimum and maximum frequencies of torsional oscillations 
in the bubble phase.  We remark in passing that in the minimal $\tilde{H}$ case, 
neutrons inside the bubbles are assumed to oscillate without escaping from the bubble 
surface or inviting neutrons outside the surface to come in.

\begin{figure*}
\begin{center}
\begin{tabular}{cc}
\includegraphics[scale=0.5]{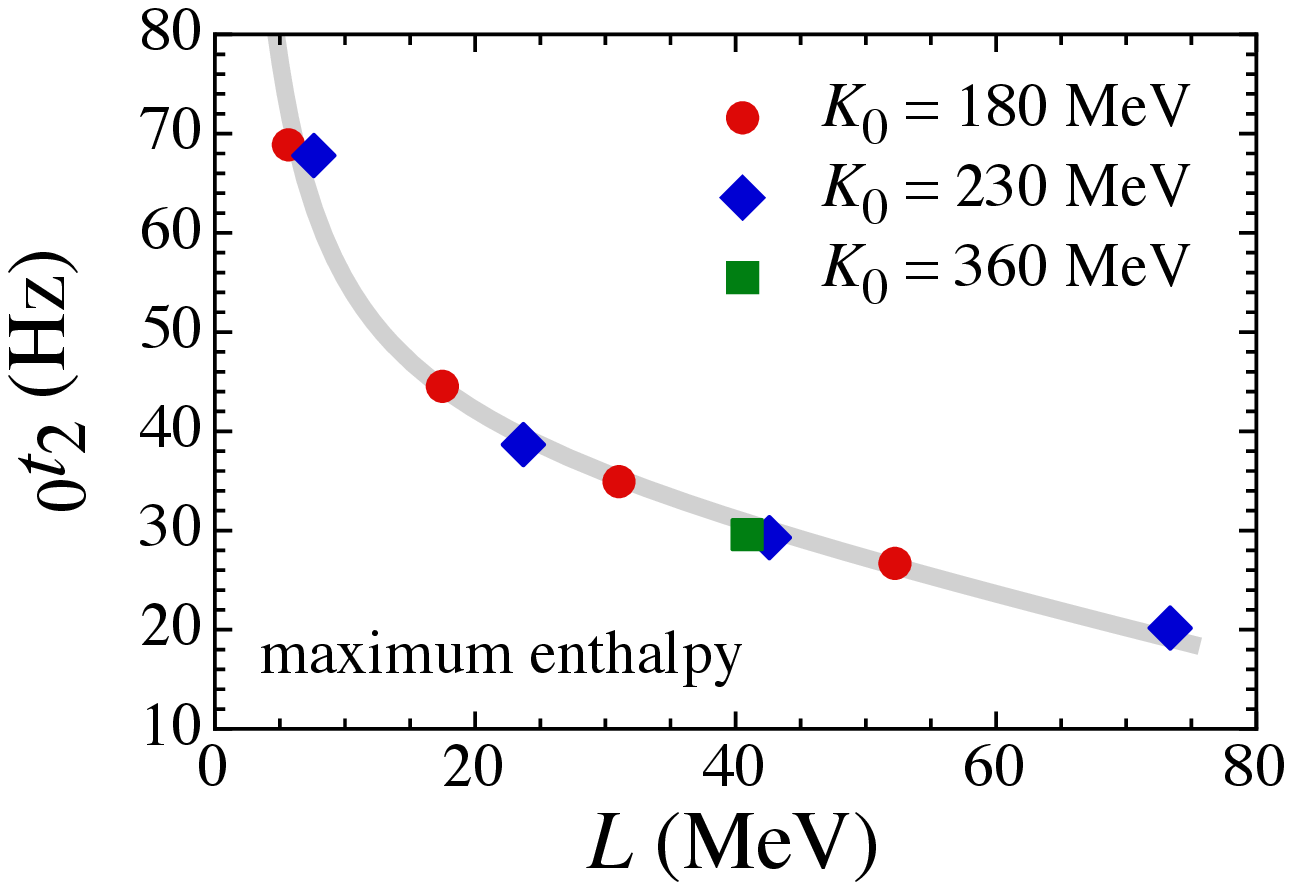} &
\includegraphics[scale=0.5]{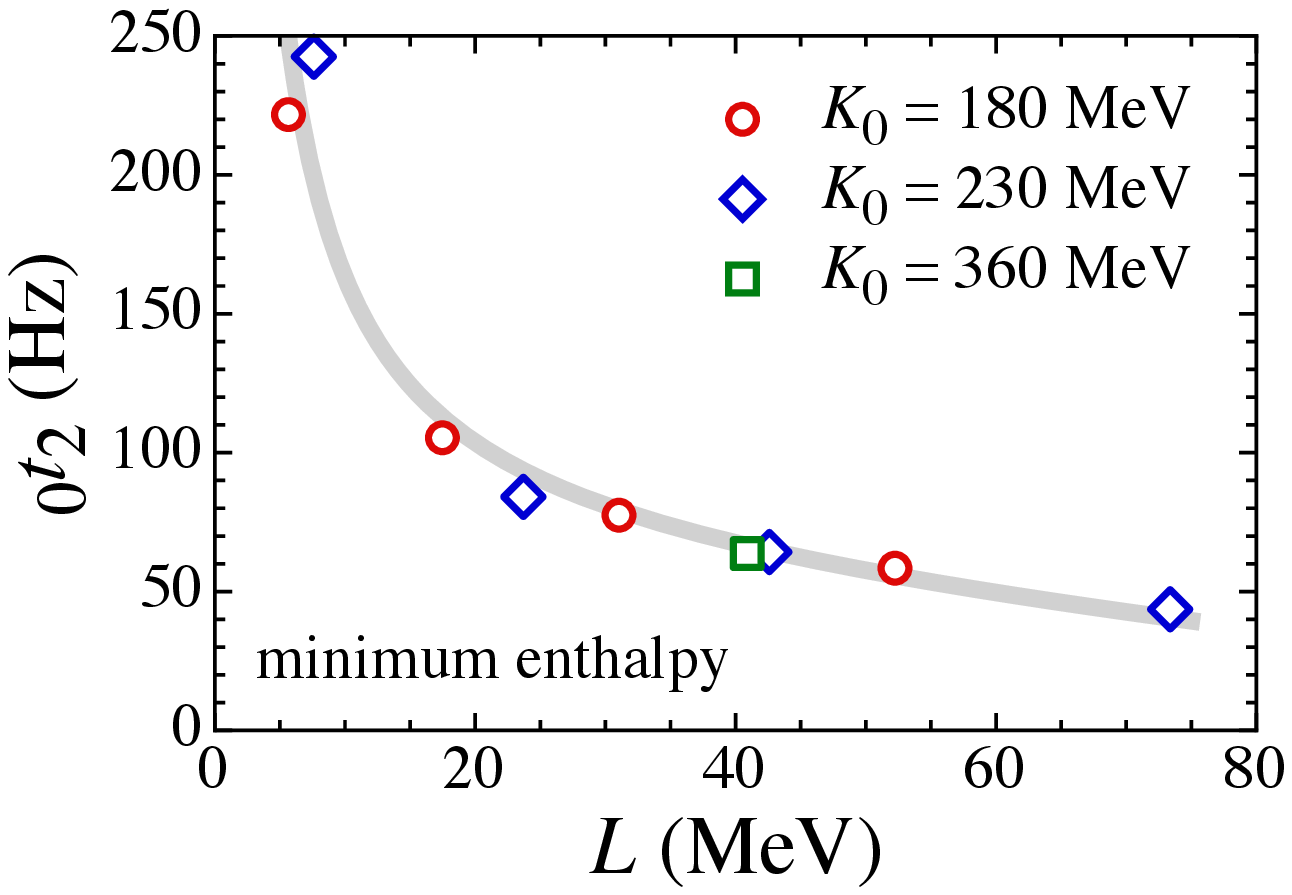}
\end{tabular}
\end{center}
\caption{
(Color online) Eigenfrequencies of the $\ell=2$ fundamental torsional oscillations in the bubble phase, 
calculated for the stellar models with $M=1.4M_\odot$ and $R=12$ km, as well as with 
various combinations of $K_0$ and $L$.  The circles, diamonds, and squares denote the 
frequencies obtained for $K_0=180$, 230, and 360 MeV, respectively.  The left panel 
corresponds to the numerical results in the case in which all the constituents 
contribute to the oscillations, while the right panel, in the case in which only 
the dripped neutrons inside a bubble contribute to the oscillations.  In both panels, the 
thick solid lines denote the fitting given by Eq.\ (\ref{eq:fitl}).
}
\label{fig:0t2-M14R12}
\end{figure*}

For the stellar models with $M=1.4M_\odot$ and $R=12$ km that are constructed from 
the EOS with various sets of $K_0$ and $L$, the $\ell=2$ fundamental frequencies of 
torsional oscillations in the bubble phase are calculated for the two extreme 
cases of the enthalpy mentioned above.  The numerical results are shown in 
Fig.\ \ref{fig:0t2-M14R12}, where the left and right panels correspond to the results 
for the maximal and minimal enthalpy that contributes to the oscillations, 
respectively.  One can observe that the frequencies of torsional oscillations in the 
bubble phase are almost independent of the values of $K_0$, but strongly depend 
on the value of $L$.  This $L$ dependence arises mainly from the fact that the 
proton density outside the bubbles and hence $Z_{\rm bubble}$ decreases with $L$.  In fact, 
the frequencies for the maximal and minimal enthalpies can be well fitted as a 
function of $L$ via 
\begin{equation}
  {}_0t_2 = d_2^{(0)} / L + d_2^{(1)} + d_2^{(2)}L, \label{eq:fitl}
\end{equation}
where $d_2^{(0)}$, $d_2^{(1)}$, and $d_2^{(2)}$ are the adjustable parameters.  The 
resultant fitting lines are also shown in Fig.\ \ref{fig:0t2-M14R12} for $L\le 75$ MeV.  
We remark that the fundamental frequencies of torsional oscillations in the crustal 
region composed of spherical nuclei can also be well fitted as a function of $L$, 
but the functional form is different from Eq.\ (\ref{eq:fitl}) \citep{SNIO2012,SNIO2013a}.  
We also remark that in the case of the minimal enthalpy (no entrainment), the 
eigenfrequencies in the bubble phase are higher than those in the phase of 
spherical nuclei by a factor that increases with decreasing $L$.  This is mainly 
because the gas density is smaller than the liquid density, while decreasing with
decreasing $L$.

\section{Comparison with the QPO frequencies}
\label{sec:IV}

\begin{figure}
\begin{center}
\includegraphics[scale=0.5]{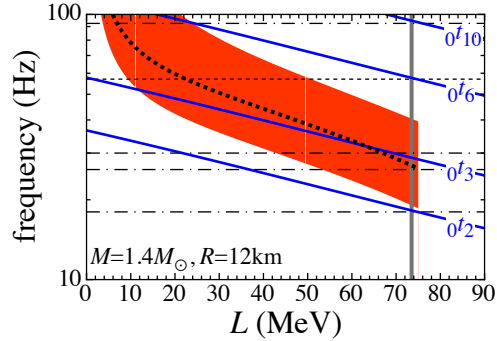}
\end{center}
\caption{
(Color online) Comparison of the observed QPO frequencies in SGR 1806$-$20 with the crustal 
torsional oscillations calculated for the stellar model with $M=1.4M_\odot$ and 
$R=12$ km.  The horizontal lines denote the observed QPO frequencies, where the dot-dash 
lines correspond to 18, 26, 30, and 92.5 Hz discovered in the giant flare, and 
the broken line corresponds to 57 Hz found from the shorter and less energetic
recurrent 30 bursts. The solid lines represent the calculated $\ell=2$, 3, 
6, and 10 fundamental frequencies of torsional oscillations in the crustal region
composed of spherical nuclei.  The vertical solid line denotes the most suitable value 
of $L$, 73.5 MeV, for explaining the observed QPOs except 26 Hz in terms of 
the crustal torsional oscillations.  The shaded area denotes the evaluated 
frequency range of the $\ell=2$ fundamental oscillation in the bubble phase, 
which covers the allowed values of the participant ratio $\tilde{H}/H$.  
The dotted line is the result for the 50 $\%$ participant ratio.
}
\label{fig:Bubble-M14R12}
\end{figure}

We proceed to compare the observed QPO frequencies from SGRs with 
the calculated frequencies of torsional oscillations in the bubble phase, 
together with those of usual torsional oscillations in the crustal region 
composed of spherical nuclei.  We have already discussed the possibility 
of identifying the observed lowest three QPOs (18, 30, 92.5 Hz) 
except the 26 Hz QPO in the giant flare of SGR 1806$-$20 as the 
$\ell=2$, 3, 10 fundamental crustal torsional oscillations \citep{SNIO2013b}.  
In Fig.\ \ref{fig:Bubble-M14R12}, we show such an identification in 
SGR 1806$-$20 for the stellar models with $M=1.4M_\odot$ and $R=12$ km.  One can 
find from this figure that the most suitable value of $L$ for explaining the 
QPOs in terms of the crustal torsional oscillations that have the 
entrainment effect included by following \cite{Chamel2012} is $L=73.5$ MeV.  
We remark that the QPO frequency (57 Hz) discovered from the shorter and less 
energetic recurrent 30 bursts \citep{QPO2} can also be identified as the 
$\ell=6$ fundamental crustal torsional oscillation with the same value of 
$L$.  With all these successful identifications, there is a problem 
with explaining the remaining QPO frequency, i.e., 26 Hz, because the interval 
between 26 and 30 Hz is too small to explain by the fundamental 
crustal torsional oscillations with neighboring $\ell$ as long as 
the lowest QPO is identified as the lowest crustal torsional mode with $\ell=2$ 
\citep{Sotani2007}.  Note that if the lowest QPO is identified as the crustal 
torsional mode with $\ell=3$, one can explain all the low-lying QPOs in terms of 
the crustal torsional modes only when the value of $L$ is assumed to be of order 
or even larger than 100 MeV \citep{SNIO2013b,SIO2016}, being significantly larger 
than the values deduced from experiments \citep{Tsang2012}.

It is thus interesting to search for the possibility of explaining 
the 26 Hz QPO in terms of torsional oscillations in the bubble phase
in the same stellar model.  Recall that these modes can coexist with crustal
torsional oscillations in the phase of spherical nuclei, thanks to vanishing
shear modulus in the slab phase, and that the $\ell=2$ fundamental frequency of 
torsional oscillations in the bubble phase is expected to lie in a range 
between the frequencies shown in Fig.\ \ref{fig:0t2-M14R12} in the cases of 
the maximal and minimal enthalpy that contributes to the torsional 
oscillations.  This range corresponds to the shaded region in Fig.\ 
\ref{fig:Bubble-M14R12}.  Interestingly, 26 Hz is in the middle of this range 
at the optimal $L$.  In fact, to identify the 26 Hz QPO as the fundamental 
torsional oscillation in the bubble phase, the participant ratio 
$\tilde{H}/H$ should be close to 50 $\%$, as can be seen from Fig.\ 
\ref{fig:Bubble-M14R12}.  Here, the participant ratio is closely related 
to the entrainment effect; to deduce what portion of nucleons outside bubbles are 
locked to the motion of the bubbles, we have only to know the participant ratio 
in the absence of the entrainment effect, which is of order 5--20 $\%$.
Note that the $L$ values at which the bubble phase is predicted to occur 
marginally contain the optimal value of 73.5 MeV.  This suggests that neutron 
stars of $M\ge1.4M_\odot$ and $R\ge12$ km are favored by the present scenario 
that requires the presence of bubbles, because the calculated eigenfrequencies 
and hence the optimal value of $L$ tend to decrease with increasing $R$ 
and/or $M$.

\begin{figure}
\begin{center}
\includegraphics[scale=0.5]{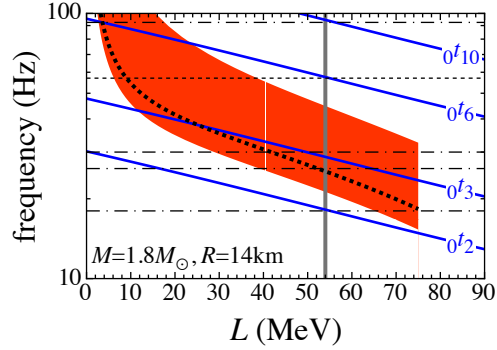}
\end{center}
\caption{
(Color online) Same as Fig.\ \ref{fig:Bubble-M14R12} for $M=1.8M_\odot$ and $R=14$ km.  
In this case, the dotted line is the result for the 70 $\%$ participant 
ratio.
}
\label{fig:Bubble-M18R14}
\end{figure}

To demonstrate that the above scenario works more reasonably for larger 
and heavier neutron stars, we have repeated the same analysis for the 
stellar model with $M=1.8M_\odot$ and $R=14$ km; the results are  
exhibited in Fig.\ \ref{fig:Bubble-M18R14}.   This figure shows that 
the same identifications of the observed QPOs work again, while the optimal 
values of $L$ and $\tilde{H}/H$ are now $\sim54$ MeV and $\sim70$ $\%$, 
respectively.  We remark in passing that the optimal participant ratio in 
the bubble phase changes with the stellar model, which may open up the
way of constraining the star's mass and radius once the entrainment effect
is known from band calculations.

\section{Conclusion}
\label{sec:V}

In summary,  we have examined torsional oscillations in the bubble 
phase located just above the crust-core boundary of neutron stars.  The corresponding 
eigenfrequencies of the fundamental modes have been calculated for various models of 
the crust EOS,  for various values of the star's mass and radius, as well as for various 
values of the participant ratio that reflects the entrainment effect, i.e., what portion 
of nucleons outside the bubbles comove with the oscillating bubbles.  The resultant 
eigenfrequencies in the bubble phase are appreciably higher than the ones in the phase 
of spherical nuclei.  This feature allows one to search for the possibility of 
reproducing the low-lying QPO frequencies observed from SGRs by appropriately 
identifying the low-lying QPOs either as a torsional oscillation in the bubble phase 
or as a usual crustal oscillation in the phase of spherical nuclei, as well as by 
keeping the value of $L$ reasonable.  By simplified calculations, we have succeeded in 
finding out such a possibility.  To make better estimates, however, many questions remain.  
It would be significant to examine the entrainment effect in the bubble phase based on 
band calculations \citep{Chamel2012}.  For completeness, possible coupling with 
propagating shear modes in the cylindrical, slab, and cylindrical-hole phases should be 
allowed for.  Magnetic fields, shell and pairing effects on bubbles, electron screening, 
polycrystalline nature, etc.\ have been also ignored, but would play a role in modifying 
the eigenfrequencies in the bubbles phase.

This work was supported in part by Grants-in-Aid for Scientific Research on Innovative Areas through No.\ 15H00843 and No.\ 24105008 provided by MEXT and by Grant-in-Aid for Young Scientists (B) through No.\ 26800133 provided by JSPS.



\end{document}